\documentstyle[multicol,epsfig,prl,aps]{revtex}

\ifpreprintsty
\def\multb{ }
\def\multe{ }
 \else
\def\multb{ \begin{multicols}{2}}
\def\multe{ \end{multicols}}
 \fi

\begin{document}
\title{Density Functional Theory of Magnetic Systems Revisited}
\author{H. Eschrig$^{\dag }$ and
       W. E. Pickett$^{\dag \dag}$ 
}
\address{$^{\dag }$Institut f\"ur Festk\"orper- und Werkstofforschung 
        Dresden e.V., Postfach 270016, D-01171 Dresden, Germany \\
$^{\dag \dag }$Department of Physics, University of California, Davis,
              CA 95616 \\
}
\date{\today}
\maketitle

\begin{abstract}
The Hohenberg-Kohn theorem of density functional theory (DFT) for
the case of electrons interacting with an external magnetic field (that
couples to spin only) is examined in more detail than previously.
An unexpected generalization is obtained: in certain cases (which include
half metallic ferromagnets and magnetic insulators) the ground state,
and hence the spin density matrix, is invariant for some non-zero range
of a shift in uniform magnetic field.
In such cases the ground state energy is not a functional of
the spin density matrix alone.  
The energy gap in an insulator or a half metal is shown
to be a ground state property of the N-electron system in magnetic DFT.
\end{abstract}

\multb
The half metallic state of a ferromagnet has been
receiving greatly increasing attention since its
prediction from band theory\cite{HM} to be the ground state of 
important magnetic
materials such as CrO$_2$,\cite{cro2} NiMnSb,\cite{nimnsb}
and Sr$_2$FeMoO$_6$\cite{doubperov} 
and several other intermetallics and oxides, and its unusual
physical properties.\cite{irkhin}  
Such systems have become
very attractive for magnetoelectronic applications, where control of
the spin degree of freedom is already leading to new devices.\cite{prinz} 
Materials thought to be half metals
have been connected with the phenomena of colossal magnetoresistance
(CMR),\cite{CMR} large tunneling MR,\cite{soulen} and 
large, low-field intergrain MR,\cite{IMR} and they would be optimal
for applications of spin valve systems\cite{nimnsb} for non-volatile
magnetic memory and for high density magnetic storage.

The half metallic state
is, in a one-electron picture, a collinear magnetic state 
in which one spin direction is metallic
while the other is gapped (`insulating').  This state is half metallic
in another sense: the absence of low energy spin flips leads to a
vanishing magnetic susceptibility like an insulator, but its charge
response (conductivity) is that of a metal.  These properties combine to give a
one-electron description of a spin-charge separated state.  In fact,
almost all understanding of half metals so far is based
on the one-electron picture, which opens up questions such as
(1) what is a half metal in many body context, and (2)
are there other unusual possibilities in magnetic systems?  One general
characterization might be in terms of conductivity (charge response)
and susceptibility (spin response) alluded to above:  
in an insulator both vanish,
in a conventional (even ferromagnetic) metal both are non-zero, and in a
half metal the conductivity is non-zero while the susceptibility
vanishes.  A clear many body formulation is however lacking.

Since density functional theory (DFT) is 
a rigorous many body theory for (chosen)
ground state properties,  
we revisit the foundations of DFT
with magnetic properties in mind.  
The first Hohenberg-Kohn (HK) theorem\cite{HK}, 
which is the basis for the 
DFT of spin-independent particles, demonstrates the
existence of a unique map
\begin{equation}
n({\bf r}) \mapsto v({\bf r})~ {\rm mod~(constant)},
\label{eq1}
\end{equation}
where $v$ is the external potential and $n$ is the ground state
particle density.  
According to the second HK theorem, the ground state energy and density
are obtained as the solution to a variational principle:
\begin{equation}
E[v-{\mu}] = min_n \{F[n] + \int n ( v-{\mu}) d^3 r\},
\end{equation}
with $\mu$ the chemical potential.
Although the variational principle has been put on an independent,
more general basis,\cite{LL} the uniqueness of the map (\ref{eq1}) remains an
important issue regarding the existence and uniqueness of the functional
derivative $\delta F/\delta n = -(v-\mu)$.
DFT, as extended by Kohn and Sham\cite{KS} and many others,
forms the basis of our understanding of the electronic behavior of
real materials.  The theory has been extended to electrons with 
spin\cite{vBH,IEG}
and also applied heavily, however the HK theorem for
interacting particles with spin has repeatedly been stated to be
analagous to the HK theorem, although this was already
questioned in \cite{vBH}.
Zero susceptibility, however, would imply that the ground state spin
density does not change when an external magnetic field is changed.
In this paper we construct a more revealing generalization of the HK theorem,
obtain explicitly the
conditions that allow half metallicity, and demonstrate some unexpected
consequences.

We consider the system in an external 
magnetic field ${\bf B}({\bf r})$ in the (commonly considered)
non-relativistic limit, in which
the field acts only on the
electron spin and the dipolar interaction between spins is 
neglected.
The potentials can be combined into 2$\times $2 spin matrix
\begin{equation}
u_{ss^{\prime }}({\bf r})=v({\bf r})\delta _{ss^{\prime }}-\mu _{B}{\bf B}(%
{\bf r})\cdot {\vec \sigma }_{ss^{\prime }}.
\end{equation}
The external field ${\bf B}$ may vary in magnitude
and direction.

Realizing that two different scalar potentials cannot lead to the
same ground state $\Psi$, the original derivation of HK\cite{HK} concluded 
that if $\Psi\mapsto v-\mu$ is unique then $n\mapsto
v-\mu$ is unique.
Following the original derivation of HK\cite{HK},
we begin by supposing that there are two
different potentials $u, u^{\prime}$ which lead to the same ground state $\Psi$.
We show that $\Psi\mapsto (v-\mu,\vec B)$ is {\it not} a unique
mapping in general.

The many-body Hamiltonian of the system is 
\begin{equation}
\hat{H}=\hat{T}+\hat{W}+\hat{U},
\label{Hamil}
\end{equation}
where $\hat{T}$ is the kinetic energy operator, $\hat{W}$ is the Coulomb
interaction energy, and $\hat{U}$ is the interaction with the external
potential. The fermionic many-particle Schr\"{o}dinger equation is 
(in atomic units)
\begin{eqnarray}
\Bigl[\sum_{i}^{N}\frac{-\nabla _{i}^{2}}{2} + \sum_{i<j}^{N}w({\bf r}%
 _{i},{\bf r}_{j}) \Bigr] \Psi ({\bf r}_{1}\alpha_{1},..,{\bf r}_N{\alpha_N})
   \\ \nonumber
 + \sum_{i}^{N}\sum_{\beta _{i}}u_{\alpha _{i},\beta _{i}}({\bf r}_{i})
  \Psi ({\bf r}%
_{1}\alpha_{1},..,{\bf r}_i\beta_i,..,{\bf r}_{N}\alpha_{N})\\ \nonumber
  =E\Psi ({\bf r}_{1}\alpha _{1},...,{\bf r}_{N} \alpha _{N}),
\end{eqnarray}
where ${\bf r}_i, \alpha_i$ are the space and spin coordinates of the
$i$-th electron;
$w({\bf r},{\bf r}^{~\prime }) = e^2/|{\bf r}-{\bf r}^{\prime}|$ is the 
Coulomb interaction.

Assume there are two external potentials $u, 
u^{\prime}$
with energies $E, E^{\prime}$ that have the same ground state
wave function $\Psi ({\bf r}_{1}\alpha _{1},...,{\bf r}_{N}\alpha _{N})$.
Subtracting the two many-particle Schr\"{o}dinger equations leads to 
\begin{eqnarray}
\sum_{i=1}^{N} \sum_{\beta_i} \Delta u_{\alpha _{i},\beta _{i}}({\bf r}_{i})
\Psi ({\bf r}_{1}\alpha_{1},..,{\bf r}_i\beta_i,..,{\bf r}%
_{N}\alpha_{N})= \\ \nonumber
\Delta E ~\Psi ({\bf r}_{1}\alpha _{1},...,{\bf r}%
_{N}\alpha _{N}),
\end{eqnarray}
where $\Delta u = u-u^{\prime}, \Delta E = E-E^{\prime}$.
Now, we perform a unitary spin rotation 
$Q_{ss^{\prime }}({\bf r})$
at each point of space that diagonalizes the difference in
potentials ({\it i.e.} rotates $\vec B$ to lie along the $\hat z$ direction:
\begin{equation}
\{Q({\bf r})[\Delta u({\bf r})]Q^{\dag }({\bf r})\}_{ss^{\prime }} =
\Delta \tilde{u}_s({\bf r}) \delta _{ss^{\prime }}.
\label{diag}
\end{equation}
The wavefunction is transformed according to 
\begin{eqnarray}
\prod_{i}^{N}Q_{\alpha _{i}\alpha _{i}^{\prime }}({\bf r}_{i})\Psi ({\bf r}%
_{1}\alpha _{1}^{\prime },...,{\bf r}_{N}\alpha _{N}^{\prime })
\equiv \tilde{%
\Psi}({\bf r}_{1}\alpha _{1},...,{\bf r}_{N}\alpha _{N}), \nonumber
\end{eqnarray}
where $\prod_{i}^{N}Q_{\alpha _{i}\alpha _{i}^{\prime }}({\bf r}_{i})$ is
the operator that rotates each of the ($\alpha _{i}$).
Collecting these results gives 
\begin{eqnarray}
\sum_{i=1}^{N}\Delta \tilde{u}_{\alpha _{i}}({\bf r}_{i})
   \tilde{\Psi}({\bf r}_{1}\alpha _{1},...,{\bf r}%
_{N}\alpha _{N})  
=\Delta E ~\tilde{\Psi}({\bf r}_{1}\alpha _{1},...,{\bf r%
}_{N}\alpha _{N}). \nonumber
\end{eqnarray}

$\tilde{\Psi}$ is some $\{{\bf r}_i\}$-dependent
multi-component function of the $2^{N}$ possible spin
configurations, at least one of which
must be non-zero. Choose a non-zero component $%
\tilde \Psi _{c}$ and denote by $N_{\uparrow }$ 
the number of $\alpha _{i}=\uparrow 
$ values in this component. Since $\tilde{\Psi}$ is antisymmetric (as was $%
\Psi $) with respect to permutations of (${\bf r}_{i}\alpha _{i}$) with ($%
{\bf r}_{j}\alpha _{j}$), we may renumber the particle indices in such a way
that $\alpha _{1}=\alpha _{2}=...=\alpha _{N_{\uparrow }}$, $\alpha
_{N_{\uparrow }+1}=\alpha _{N_{\uparrow }+2}=...=\alpha _{N}.$ This ordering
lets us write 
\begin{eqnarray}
\left\{ \sum_{i=1}^{N_{\uparrow}} \Delta \tilde{u}_{\uparrow }({\bf r}_{i})
+\sum_{i=N_{\uparrow}+1}^N \Delta \tilde{u}_{\downarrow }({\bf r}_{i})
\right\} \tilde{\Psi_c}({\bf r}%
_{1}\uparrow ,...,{\bf r}_{N}\downarrow ) \\ \nonumber
=\Delta E \tilde{\Psi_c}({\bf r}%
_{1}\uparrow ,...,{\bf r}_{N}\downarrow ).  
\label{spinsplit}
\end{eqnarray}
This equation must hold for all values of (${\bf r}_{1},...,{\bf r}_{N})$.
(We suppose $u, u^{\prime}$ are analytic in ${\bf r}$
except possibly at isolated points,
so that  $\tilde{\Psi_c}$ is non-zero almost everywhere.)
By varying only ${\bf r}_1$, and then separately varying only ${\bf r}_N$,
we obtain 
\begin{equation}
\Delta \tilde u_{\uparrow} = C_{\uparrow},~~
\Delta \tilde u_{\downarrow} = C_{\downarrow},
\end{equation}
where $C_{\uparrow}, C_{\downarrow}$ are constants.
The special cases $N_{\uparrow}=0$ or $N_{\uparrow}=N$ do not lead
to new consequences.
For further analysis, we consider separate cases.

\noindent {\it Case A: impure spin states.}
Suppose that there are at least two components of $\tilde{\Psi}$ with
different values of $N_{\uparrow }$ and hence $N_{\downarrow }=N-N_{\uparrow
}$.  Then
\begin{equation}
N_{\uparrow }C_{\uparrow }+(N-N_{\uparrow })C_{\downarrow }=\Delta E.
\end{equation}
Since this holds for two different values of $N_{\uparrow }$, it follows
that 
$C_{\uparrow }=C_{\downarrow } \equiv C,$
which leads to $\Delta \tilde u_{\uparrow}=\Delta \tilde u_{\downarrow}$
so $\Delta u_{\alpha,\beta}=C\delta_{\alpha,\beta}$.
By the ground state energy minimum principle,
this recovers the usual Hohenberg-Kohn result 
\begin{eqnarray}
n_{ss^{\prime}}=n^{\prime}_{ss^{\prime}} \rightarrow \left( 
\begin{array}{c}
v({\bf r}) - v^{\prime}({\bf r}) \equiv C, \\ 
{\bf B}({\bf r}) - {\bf B}^{\prime}({\bf r}) \equiv 0,
\end{array}
\right)  \label{conditions}
\end{eqnarray}
implying a non-zero susceptibility.

\noindent {\it Case B: pure spin states.}
Suppose now that all non-zero components of 
$\tilde{\Psi}$ have the same value
of $N_{\uparrow }$ and $N_{\downarrow }$. These may be considered as ``pure
spin'' states, eigenfunctions of the operator $\hat {S}^z$ = 
$\sum_i \sigma^z_{\alpha_i \beta_i}/2$ with eigenvalues
$S_z=N_{\uparrow }-(N/2)=(N_{\uparrow }-
N_{\downarrow })/2$.
Then $C_{\uparrow }$ and $C_{\downarrow }$ need not be equal and we can
write 
\begin{eqnarray}
\Delta \tilde{u}& = & \left( 
  \begin{array}{cc}
  C_{\uparrow } & 0 \\ 
  0 & C_{\downarrow }
  \end{array} 
\right)      
=\bar C {\bf 1} - \mu_B \bar{B}\sigma^{z},
\end{eqnarray}
where $\bar{C} = (C_{\uparrow} + C_{\downarrow}$)/2 and 
$-\mu_B \bar{B} = (C_{\uparrow} - C_{\downarrow}$)/2.

Backtransforming according to the inverse of Eq. (\ref{diag}) gives 
\begin{equation}
 \Delta u_{\alpha \beta }({\bf r})=\bar C \delta _{\alpha
\beta }+\mu_B \bar{B}[Q^{\dag }({\bf r})\sigma _{z}Q({\bf r})]_{\alpha \beta }.
\end{equation}
The last term on the right is position-dependent, non-diagonal, and
non-vanishing in general. In this case the conditions for identical
ground state wavefunctions are
\begin{equation}
\Psi =\Psi ^{\prime }\rightarrow
\left(
\begin{array}{ccc}
v({\bf r})-v^{\prime }({\bf r}) & = & \bar{C} \hfill \\
{\bf B}({\bf r})-{\bf B}^{\prime }({\bf r}) & = & \bar{B} ~ \hat e({\bf r}).
\end{array}
\right) .
\label{condition3}
\end{equation}
where $\hat e$ is the unit vector $\frac{1}{2}Tr\{\vec{\sigma}
Q^{\dag }({\bf r})\sigma _{z}Q({\bf r})\}$.  The result (\ref{conditions})
is modified accordingly.
This result is a highly non-trivial generalization of the HK theorem:
{\it two magnetic fields
whose difference is constant in magnitude, but possibly is non-unidirectional,
may give rise to the same ground state.}

Now we investigate the conditions on $Q$ for which $\tilde \Psi$ is an 
eigenstate of $\hat {S}^z$, {\it i.e.} $\tilde \Psi$ describes a collinear
spin arrangement.  
Considering the Hamiltonian 
Eq. (\ref{Hamil}), there must be an operator 
\begin{equation}
\hat U_o = \sum_{i=1}^N Q^*_{\alpha^{\prime}_i \alpha_i}
({\bf r}_i) \sigma^z_{\alpha^{\prime}_i \beta^{\prime}_i}
Q_{\beta^{\prime}_i \beta_i}({\bf r}_i)
\end{equation}
that commutes with $\hat T + \hat U + \hat W$.  We now specialize
to the particular case where
one of the external fields, $\bf B^{\prime}$, is zero.  
Since the interaction $\hat W$ is spin
independent,  $\hat U_o$ will commute with $H^{\prime}$ 
if and only if it commutes 
with $\hat T$.
One can show that this necessitates that
$Q$ be ${\bf r}$-independent, so that $\Psi$ itself is an eigenstate of
$\hat {S}^z$, and hence is a collinear spin state.  The second condition in Eq.
(\ref{condition3}) reduces to ${\bf B} - {\bf B^{\prime}} = B \hat z$:
a turning on of a uniform
magnetic field leaves the ground state invariant. Restated: in the
subspace of collinear magnetizations, {\it the ground state determines the
magnetic field only up to some codirectional uniform field}.  
A direct corollary 
is that there is no longer any ground state
energy functional $E[n]$ of the density $n_{ss'}(\bf r)$ alone
(see Eq. (\ref{eq18}) below).

In the collinear situation, inserting $\Delta u = \mu_B B\sigma_z$
into Eq. (6) yields 
\begin{equation}
 \Delta E(B) = -(N_\uparrow-N_\downarrow) \mu_B B, 
\label{eq18}
\end{equation}
which gives the well known dependence of energy vs. field for a
system of fixed spin.
Consider as a simple example a Be atom in a uniform magnetic field,
with its ground state characterized as
$1s^22s^2$ ($N=4$, $N_\uparrow=2$). The lowest excited
$\hat S_z$-eigenstate is $1s^22s2p$ with $N_\uparrow=3$. Its excitation energy 
is that of a $2s\rightarrow 2p$ promotion. There is another excited state
$1s2s^22p$ with the same $N_\uparrow$, but the much higher excitation energy
of a $1s\rightarrow 2p$ core excitation. The energetically lowest
$N_\uparrow=4$ state is $1s2s2p^2$ whose excitation energy is roughly the sum
of the previous two. The situation is sketched in Fig. 1(a),
where the lines with positive slopes correspond to 
states with all spins reversed.

Since states with $N_\uparrow=N/2\pm n$ are degenerate for $B=0$,
Fig. 1(a) may be supplemented symmetrically to the vertical axis. Hence, for
$|B| < B_0$ the groundstate is $1s^22s^2$ with energy
$E_0$, for $B_0 < B < B_1$ the ground state is $1s^22s2p$ with
energy $E_1-2\mu_B B$, and for $B\ge B_1$ the ground state is
$1s2s2p^2$ with energy $E_3-4\mu_B B$.  The ground state does not change
with field except at certain isolated values.

In an extended system, say a non-magnetic insulator with gap $\Delta_g$, 
there is a continuum above $\Delta_g$ (one excited electron
with reversed spin), another continuum above $2\Delta_g$ (two excited electrons) 
and so on, as illustrated in Fig. 1(b).
In an extended system one would prefer to consider the intensive quantity
\begin{equation}
 \frac{\Delta E(B)}{N} 
= -\mu_B B \left(\frac{N_\uparrow - N_\downarrow}{N}\right)
\end{equation}
instead of $\Delta E$ itself. Then, one finds that for $\mu_B B < 
\Delta_g$ the groundstate is independent of
$B$, beyond which the state changes and $\Delta E/N$ 
veers off. 
Thus while the gap $\Delta_g$ is not a 
ground state property of the $N$ particle system in paramagnetic DFT
(it involves the N$\pm$1 particle ground states),
it is a ground state property in the presence of
a uniform field.

For a stoichiometric half metal with moment per cell 
$\mu_B {\cal M}~({\cal M}$ an
integer) the picture is related, except there is an overall bias -- a
slope of -$\mu_B {\cal M}$ in the energy per cell -- and the 
positive and negative $B$
directions are not symmetric.
The situation that is sketched in Fig. 1(c) has a gap
$\Delta_v+\Delta_c$ for $\downarrow$ spin states, with no gap for 
$\uparrow$ spin.
The chemical potential $\mu$ corresponds to the energy to remove an
$\uparrow$ spin, and the quantities $\Delta_v = \mu_B B_v$, 
$\Delta_c = \mu_B |B_c|$ represent the energy, or field, required to flip
a spin from $\downarrow$ to $\uparrow$, or vice versa.
Note again that the interval of $B$ for which the state does not 
change, which is the gap in the $\downarrow$ spectrum, is a ground
state property of the $N$ particle system in an external magnetic field.

It is useful to consider the form of
constrained DFT in which $N_{\uparrow}$ and $N_{\downarrow}$ are specified,
which leads to two associated chemical potentials $\mu_{\uparrow}$,
$\mu_{\downarrow}$.  Then as $N_s$ is changed to $N_s \pm 1$, $\mu_s$
may vary only to order 1/$N_s$ (metallic behavior) 
or it may jump discontinuously
across a gap, just as is the case for insulators.\cite{1983}
The half metal is defined as that situation in which one and only one of
$\mu_s$ (we have choosen $\downarrow$) is discontinuous upon 
addition of one electron.  
For an insulator, there is a discontinuity
in $\mu$ for both spins.

We now consider the KS eigenvalue spectrum.
As long as the external field
shifts the bands sufficiently little not to disturb
the half metallicity ($B_c < B < B_v$), the ground state, and hence the charge 
density in each spin channel, remain
unchanged. 
Using the same arguments as
were applied to establish the discontinuity in $v_{xc}(N(\mu))$ for an insulator
as $\mu$ crosses the gap (the kinetic energy is discontinuous across the
gap)\cite{1983}, one finds that there is a discontinuity 
in $v_{xc,\downarrow}(N_{\uparrow},N_{\downarrow})$ if the filling with 
$N_{\downarrow}$ moves $\mu_{\downarrow}$ across the gap.\cite{define} 

The Kohn-Sham gap $\varepsilon_{g\downarrow}$ is smaller than the
true (quasiparticle) gap $\Delta_g = \Delta_c + \Delta_v$.  
When the magnetic field is large enough
that $\mu$ reaches the KS band minimum $\varepsilon_c(N_{\downarrow})$, 
the occupation of that channel
becomes N$_{\downarrow}+\epsilon$
(with $\epsilon \rightarrow 0$).  This is the point of the discontinuity,
where the KS conduction eigenvalue (in fact, the entire $\downarrow$
spectrum) jumps
upward.  By comparison with Dyson's equation, and the fact that the
system's ground state spin densities must be the same whether 
obtained from DFT or the quasiparticle Greens function, this jump must
be such as to make $\varepsilon_{c\downarrow}(N_{\uparrow},%
N_{\downarrow}+\epsilon) \equiv$ $\Delta_c$,
the quasiparticle conduction band edge, for $\epsilon \rightarrow$0.

It is apparent then that the KS gap in the insulating channel is not 
equal to the true gap in that channel, and that 
$\varepsilon_c(N_{\uparrow},N_{\downarrow}) - \mu$ is not the true
spin flip energy (which is $\Delta_c$ - $\mu$).  By our
definition,\cite{define} as the reverse
field is applied and $\mu$ is driven toward the valence band maximum
$\varepsilon_{v\downarrow}$, there is no discontinuity,
and the other spin flip energy -- a true excitation energy --
is given correctly by DFT.   Needless to say, an approximation such
as the local density approximation that interpolates 
across the discontinuity, will fail
to predict both $\Delta_c$ and $\Delta_v$.

We now summarize.
We have presented new, rigorous results for the Hohenberg-Kohn mapping in
a magnetic field.  We obtain conditions that characterize half metals:
(1) two collinear systems in different uniform magnetic fields
may have the same half metallic (or magnetic insulating) ground state; 
(2) exactly one of the chemical potentials $\mu_s$ is discontinuous
upon particle addition to a half metal.
We have pointed out other
consequences, primary among them being that the ground state energy
of a system is no longer a unique functional of the density $n_{ss'}$
when magnetic fields are allowed (although the ground state itself is), 
and that the gap in a half metal is
a ground state property of the $N$ particle system.  These
results are only exact in the non-relativistic ($c \rightarrow \infty$)
limit.  For $c$ finite, half metallicity is an approximate
notion due to orbital currents and orbital moments and spin-orbit
coupling that mixes them, and the general theory\cite{vignale} 
probably restores
the conventional theorems of DFT.  Still, the notion of half metallicity will
be an important model limit.

We acknowledge stimulating discussions with I. I. Mazin throughout the 
course of this work, and helpful interaction with W. Kohn 
during the early stages.
This study was begun when the authors were at the Institute of
Theoretical Physics at the University of California at Santa Barbara,
which is supported by National Science Foundation Grant PHY-9407194. W.E.P.
was supported by National Science Foundation Grant DMR-9802076.  


\multe
\begin{figure}[tbp]
\centerline{\epsfig{file=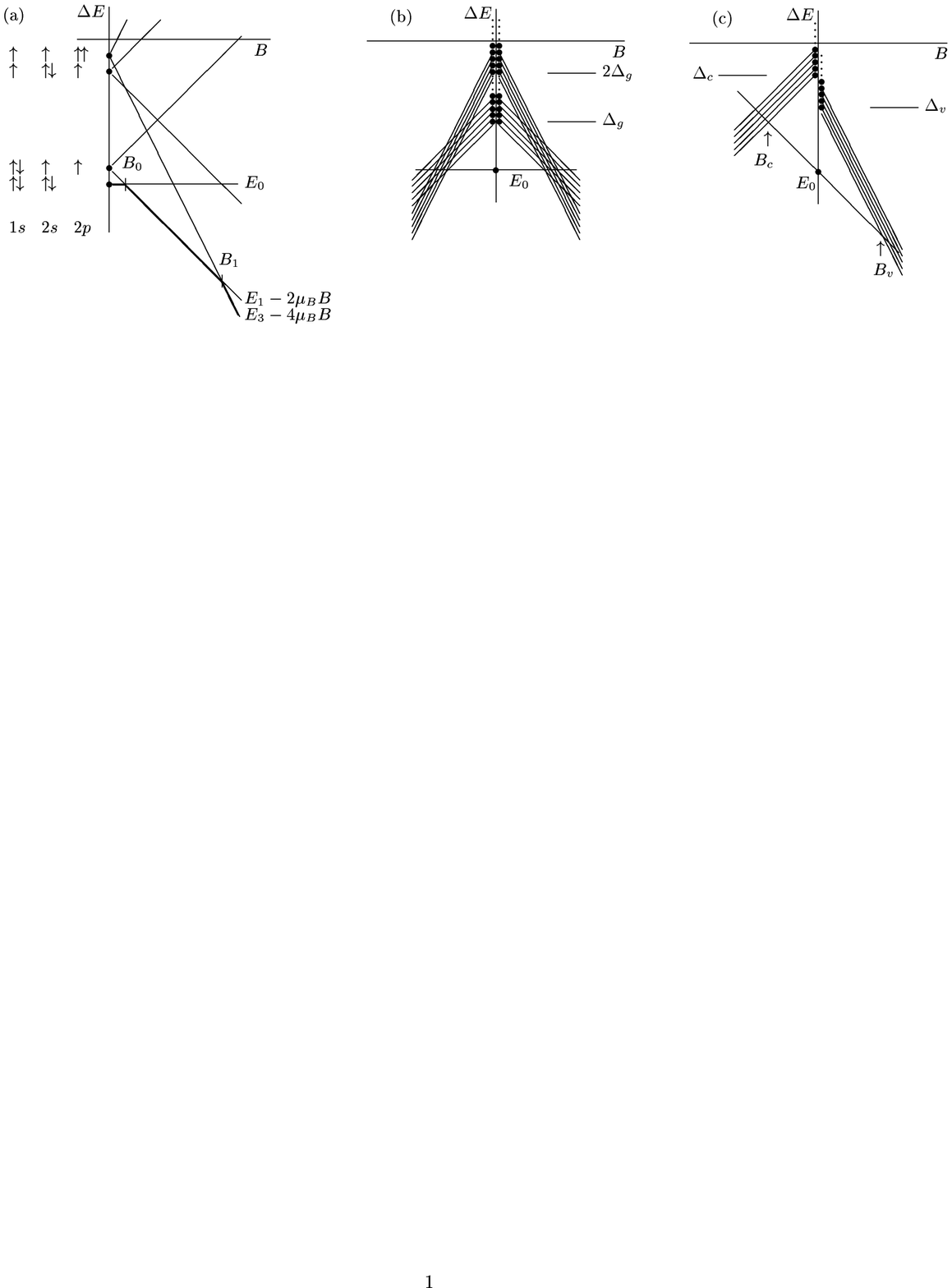,width=0.90\linewidth,angle=-00} }
\caption{Energy change with field $B$ for: (a) Be atom.  The levels are
discrete and the variation is $-S_z \mu_B B$.
(b) a nonmagnetic insulator.  The ground state energy is constant over a
range of fields $-\Delta_g < \mu_B B < \Delta_g$, beyond which 
magnetization is induced.
(c) a half metal.  The ground state energy varies as $-\mu_B (N_{\uparrow}
- N_{\downarrow}) B$ in the range $B_c < B < B_v$ 
($\Delta_v < \mu_B B <
\Delta_c$).  In (b) and (c), the small dots indicate a  continuum
extending upwards.  See text for further explanation.
}
\end{figure}


\begin{references}
\bibitem{define}We define $v_{xc\downarrow}(N_{\uparrow},N_{\downarrow})$
to be left semicontinuous at $N_{\downarrow}(\mu = \varepsilon_v)$.
\end{references}
\end{document}